\documentstyle[12pt]{article}
\def\1{{\chi}}

\begin{document}
\title {{Spectral representation of infimum of bounded quantum observables}\thanks{This project is supported by Natural Science
Found of China (10771191 and 10471124).}}
\author {Jun Shen$^{1,2}$, Jun-De Wu$^{1}$\date{}\thanks{E-mail: wjd@zju.edu.cn}}
\maketitle $^1${\small\it Department of Mathematics, Zhejiang
University, Hangzhou 310027, P. R. China}

$^2${\small\it Department of Mathematics, Anhui Normal University,
Wuhu 241003, P. R. China}

\begin{abstract} {\noindent In 2006, Gudder introduced a logic order on bounded quantum observable set $S(H)$. In 2007,
Pulmannova and Vincekova proved that for each subset $\cal D$ of
$S(H)$, the infimum of $\cal D$ exists with respect to this logic
order. In this paper, we present the spectral representation for the
infimum of $\cal D$.}
\end{abstract}


\vskip0.6in

{\bf 1. Introduction}

\vskip0.2in

Let $H$ be a complex Hilbert space, $S(H)$ be the set of all bounded
linear self-adjoint operators on $H$, $S^{+}(H)$ be the set of all
positive operators in $S(H)$, $P(H)$ be the set of all projections
on $H$. Each element in $P(H)$ is said to be a {\it quantum event},
each element in $S(H)$ is said to be a {\it bounded quantum
observable} on $H$. For $A\in S(H)$, $P^{A}$ denote the spectral
measure of $A$. Let ${\mathbf{R}}$ be the set of real numbers,
$\mathcal{B}({\mathbf{R}})$ be the set of all Borel subsets of
${\mathbf{R}}$.

Let $A, B\in S(H)$. If for each $x\in H$, $\langle Ax, x\rangle\leq
\langle Bx, x\rangle$, then we say that $A\leq B$. Equivalently,
there exists a $C\in S^{+}(H)$ such that $A+C=B$. $\leq $ is a
partial order on $S(H)$. The physical meaning of $A\leq B$ is that
the expectation of $A$ is not greater than the expectation of $B$
for each state of the system. So the order $\leq $ is said to be a
{\it numerical order} on $S(H)$.

In 2006, Gudder introduced the order $\preceq$ on $S(H)$: If there
exists a $C\in S(H)$ such that $AC=0$ and $A+C=B$, then we say that
$A\preceq B$ ([1]).

Equivalently, $A\preceq B$ if and only if for each
$0\notin\Delta\subseteq\mathcal{B}({\mathbf{R}})$,
$P^{A}(\Delta)\leq P^{B}(\Delta)$. The physical meaning of $A\preceq
B$ is that for each
$0\notin\Delta\subseteq\mathcal{B}({\mathbf{R}})$, the quantum event
$P^{A}(\Delta)$ implies the quantum event $P^{B}(\Delta)$. Thus, the
order $\preceq $ is said to be a {\it logic order} on $S(H)$ ([1]).

Let $\{A_\alpha\}_\alpha\subseteq S(H)$ be a family of bounded
linear self-adjoint operators on $H$, if there exists a $C\in S(H)$
such that $C\preceq A_\alpha$ for each $\alpha$, and $D\preceq C$
for any $D\in S(H)$ satisfies $D\preceq A_\alpha$ for each $\alpha$,
then $C$ is said to be an {\it infimum} of $\{A_\alpha\}_\alpha$
with respect to the logic order $\preceq $, and we denote
$C=\bigwedge\limits_\alpha A_\alpha$.

If $P, Q\in P(H)$, then $P\leq Q$ if and only if $P\preceq Q$, and
$P$ and $Q$ have the same infimum with respect to the orders $\leq $
and $\preceq $ ([1]).

For a given order of the bounded quantum observables set $S(H)$, the
infimum problem is to find out under what condition the infimum
$A\wedge B$ exists for $A, B\in S(H)$ with respect to the given
order? Moreover, can we give out the structure of $A\wedge B$?

For the numerical order $\leq $ of $S(H)$, the problem has been
studied in different content by Kadison, Gudder, Moreland, Ando, Du,
etc ([2-6]).

In 2007, Pulmannova and Vincekova proved that for each subset $\cal
D$ of $S(H)$, its infimum exists with respect to the logic order
$\preceq $. Their proof is abstract and there is no information
about the structure of the infimum ([7]).

In 2008, Liu and Wu found a representation of the infimum $A\wedge
B$ for $A, B\in S(H)$, but the representation is still complicated
and implicit, in particular, the spectral representation of the
infimum $A\wedge B$ is still unknown ([8]).

In this note, we present a spectral representation of the infimum
$\bigwedge\limits_\alpha A_\alpha$ for any subset
$\{A_\alpha\}_\alpha$ of $S(H)$. Our approach and results are very
different from [8] and are much more simple and explicit.

\vskip0.4in

{\bf 2. The spectral representation theorem}

\vskip0.2in

Now, we present the spectral representation theorem of infimum by
the following construction:

For each $\Delta\in \mathcal{B}({\mathbf{R}})$, if
$\Delta=\bigcup\limits^{n}_{i=1} \Delta_i$, where $\{\Delta_i\}$ are
pairwise disjoint Borel subsets of ${\mathbf{R}}$, then we say
$\gamma=\{\Delta_i\}$ is a {\it partition} of $\Delta$. We denote
all the partitions of $\Delta$ by $\Gamma(\Delta)$.

Let $\{A_\alpha\}_\alpha\subseteq S(H)$ be a family of bounded
linear self-adjoint operators on $H$.

Define $G(\emptyset)=0$. For each nonempty $\Delta\in
\mathcal{B}({\mathbf{R}})$,

\vskip0.1in

(1) if $0\notin \Delta$, define
$G(\Delta)=\bigwedge\limits_{\gamma\in \Gamma(\Delta)}
\sum\limits_{\Delta'\in \gamma}\Big(\bigwedge\limits_\alpha
P^{A_\alpha}(\Delta')\Big)$;

\vskip0.1in

(2) if $0\in \Delta$, define $G(\Delta)=I-G({\mathbf{R}}\backslash
\Delta)$.

It is clear that for each $\Delta\in \mathcal{B}({\mathbf{R}})$,
$G(\Delta)$ is a projection, so $G$ defines a map from $
\mathcal{B}({\mathbf{R}})$ to $P(H)$.

\vskip0.1in

{\bf Theorem 1.} $G: {\mathcal{B}}({\mathbf{R}})\rightarrow P(H)$ is
a spectral measure.

{\bf Proof.} (1) $G(\emptyset)=0$,
$G({\mathbf{R}})=I-G({\mathbf{R}}\backslash
{\mathbf{R}})=I-G(\emptyset)=I$.

(2) Suppose $\Delta_1,\Delta_2\in \mathcal{B}({\mathbf{R}})$,
$\Delta_1\cap \Delta_2=\emptyset$, $\Delta_3=\Delta_1\cup\Delta_2$.
There are two possibilities.

(i) $0\notin \Delta_3$.

Note that $G(\Delta_1)\leq P^{A_\alpha}(\Delta_1)$, $G(\Delta_2)\leq
P^{A_\alpha}(\Delta_2)$ and
$P^{A_\alpha}(\Delta_1)P^{A_\alpha}(\Delta_2)=0$ for any $\alpha$,
we have $G(\Delta_1)G(\Delta_2)=0$. Now, we prove that $G(\Delta_3)=
G(\Delta_1)+G(\Delta_2)$.

Let $\gamma\in \Gamma(\Delta_3)$. It is easy to see that for each
$\Delta'\in \gamma$,
$$ \bigwedge\limits_\alpha
P^{A_\alpha}(\Delta')\geq \bigwedge\limits_\alpha
P^{A_\alpha}(\Delta'\cap \Delta_1)+ \bigwedge\limits_\alpha
P^{A_\alpha}(\Delta'\cap \Delta_2)\ .$$

Let $\gamma_1=\{\Delta'\cap \Delta_1|\ \Delta'\in \gamma\}$,
$\gamma_2=\{\Delta'\cap \Delta_2|\ \Delta'\in \gamma\}$. Then
$\gamma_1\in \Gamma(\Delta_1)$, $\gamma_2\in \Gamma(\Delta_2)$, thus
$$\sum\limits_{\Delta'\in \gamma}\Big(\bigwedge\limits_\alpha
P^{A_\alpha}(\Delta')\Big)\geq \sum\limits_{\Delta'\in
\gamma}\Big(\bigwedge\limits_\alpha P^{A_\alpha}(\Delta'\cap
\Delta_1)\Big)+\sum\limits_{\Delta'\in
\gamma}\Big(\bigwedge\limits_\alpha P^{A_\alpha}(\Delta'\cap
\Delta_2)\Big)\geq G(\Delta_1)+G(\Delta_2)\ .$$

We conclude that $G(\Delta_3)\geq G(\Delta_1)+G(\Delta_2)\ .$

Now we prove the converse inequality.

Let $x\in H$ such that $G(\Delta_3)x=x$. Then for each $\gamma\in
\Gamma(\Delta_3)$, $\sum\limits_{\Delta'\in
\gamma}\Big(\bigwedge\limits_\alpha P^{A_\alpha}(\Delta')\Big)x=x$.

Let $\gamma_1\in \Gamma(\Delta_1)$, $\gamma_2\in \Gamma(\Delta_2)$.

Then $\gamma_3=\gamma_1\cup\gamma_2\in \Gamma(\Delta_3)$. Thus
$\sum\limits_{\Delta'\in \gamma_3}\Big(\bigwedge\limits_\alpha
P^{A_\alpha}(\Delta')\Big)x=x$. That is,
$$x=\sum\limits_{\Delta'_1\in \gamma_1}\Big(\bigwedge\limits_\alpha
P^{A_\alpha}(\Delta'_1)\Big)x+\sum\limits_{\Delta'_2\in
\gamma_2}\Big(\bigwedge\limits_\alpha P^{A_\alpha}(\Delta'_2)\Big)x\
.$$

Thus $$\Big(\bigwedge\limits_\alpha
P^{A_\alpha}(\Delta_1)\Big)x=\Big(\bigwedge\limits_\alpha
P^{A_\alpha}(\Delta_1)\Big)(\sum\limits_{\Delta'_1\in
\gamma_1}\Big(\bigwedge\limits_\alpha
P^{A_\alpha}(\Delta'_1)\Big)x+\sum\limits_{\Delta'_2\in
\gamma_2}\Big(\bigwedge\limits_\alpha P^{A_\alpha}(\Delta'_2)\Big)x\
)$$$$=\sum\limits_{\Delta'_1\in
\gamma_1}\Big(\bigwedge\limits_\alpha
P^{A_\alpha}(\Delta'_1)\Big)x,$$

and
$$\Big(\bigwedge\limits_\alpha P^{A_\alpha}(\Delta_2)\Big)x=\sum\limits_{\Delta'_2\in
\gamma_2}\Big(\bigwedge\limits_\alpha
P^{A_\alpha}(\Delta'_2)\Big)x.$$

We conclude that
$$\Big(\bigwedge\limits_\alpha
P^{A_\alpha}(\Delta_1)\Big)x=G(\Delta_1)x,\
\Big(\bigwedge\limits_\alpha
P^{A_\alpha}(\Delta_2)\Big)x=G(\Delta_2)x\ .$$

So $x=G(\Delta_1)x+G(\Delta_2)x$. Thus we have $G(\Delta_3)\leq
G(\Delta_1)+G(\Delta_2)\ .$ From the above we have
$G(\Delta_3)=G(\Delta_1)+G(\Delta_2)\ .$

(ii) If $0\in \Delta_3$, suppose $0\in \Delta_1$ and $0\notin
\Delta_2$.

By (i) we have $G(\Delta_2)+G({\mathbf{R}}\backslash \Delta_3)=
G({\mathbf{R}}\backslash \Delta_1)$.

Since $G(\Delta_1)=I-G({\mathbf{R}}\backslash \Delta_1)$ and
$G(\Delta_2)\leq G({\mathbf{R}}\backslash \Delta_1)$, we have
$G(\Delta_1)G(\Delta_2)=0$.

$G(\Delta_3)=I-G({\mathbf{R}}\backslash
\Delta_3)=I-\Big(G({\mathbf{R}}\backslash
\Delta_1)-G(\Delta_2)\Big)=I-G({\mathbf{R}}\backslash
\Delta_1)+G(\Delta_2)=G(\Delta_1)+G(\Delta_2)\ .$

(3) Let $\{\Delta_n\}_{n=1}^\infty\subseteq
\mathcal{B}({\mathbf{R}})$ be pairwise disjoint, we now prove that
$G(\bigcup\limits_{n=1}^\infty \Delta_n)=\sum\limits_{n=1}^\infty
G(\Delta_n)$ in the strong operator topology. By (2), we only need
to show that $G(\bigcup\limits_{i=n+1}^\infty
\Delta_i)\longrightarrow 0$ in the strong operator topology.

Since $\{\Delta_n\}_{n=1}^\infty$ are pairwise disjoint, there
exists some positive integer $N$ such that $0\notin
\bigcup\limits_{i=N+1}^\infty \Delta_i$. Thus for $n\geq N$,
$$G(\bigcup\limits_{i=n+1}^\infty
\Delta_i)=\bigwedge\limits_{\gamma\in
\Gamma(\bigcup\limits_{i=n+1}^\infty \Delta_i)}
\sum\limits_{\Delta'\in \gamma}\Big(\bigwedge\limits_\alpha
P^{A_\alpha}(\Delta')\Big)\leq \bigwedge\limits_\alpha
P^{A_\alpha}(\bigcup\limits_{i=n+1}^\infty \Delta_i)\longrightarrow
0\ .$$

By (1),(2) and (3), we completed the proof.

\vskip0.1in

{\bf Theorem 2.} $\bigwedge\limits_\alpha
A_\alpha=\int_{\mathbf{R}}\lambda dG$.

{\bf Proof.} Let $C=\int_{\mathbf{R}}\lambda dG$. Note that for each
$\alpha$, $$G\Big((-\infty,-\|A_\alpha\|)\cup(\|A_\alpha\|,+\infty)
\Big)\leq
P^{A_\alpha}\Big((-\infty,-\|A_\alpha\|)\cup(\|A_\alpha\|,+\infty)
\Big)=0\ ,$$  thus $C$ is bounded and $C\in S(H)$, in particular,
$P^C=G$.

For each $\Delta\in \mathcal{B}({\mathbf{R}})$, if $0\notin \Delta$,
we have
$$P^C(\Delta)=G(\Delta)=\bigwedge\limits_{\gamma\in
\Gamma(\Delta)} \sum\limits_{\Delta'\in
\gamma}\Big(\bigwedge\limits_\alpha P^{A_\alpha}(\Delta')\Big)\leq
\bigwedge\limits_\alpha P^{A_\alpha}(\Delta)\ .$$ Thus $C\preceq
A_\alpha$ for each $\alpha$ ([1]).

If $D\preceq A_\alpha$ for each $\alpha$, then for each
$0\notin\Delta\in \mathcal{B}({\mathbf{R}})$ and each $\gamma\in
\Gamma(\Delta)$, since $P^D(\Delta')\leq \bigwedge\limits_\alpha
P^{A_\alpha}(\Delta')$ for each $\Delta'\in \gamma$, we have
$$P^D(\Delta)=\sum\limits_{\Delta'\in
\gamma}P^D(\Delta')\leq\sum\limits_{\Delta'\in
\gamma}\Big(\bigwedge\limits_\alpha P^{A_\alpha}(\Delta')\Big)\ .$$
So we conclude that
$$P^D(\Delta)\leq\bigwedge\limits_{\gamma\in
\Gamma(\Delta)} \sum\limits_{\Delta'\in
\gamma}\Big(\bigwedge\limits_\alpha
P^{A_\alpha}(\Delta')\Big)=G(\Delta)=P^C(\Delta).$$ Thus $D\preceq
C$, so we have $\bigwedge\limits_\alpha A_\alpha=C$.

\vskip 0.2 in

\vskip0.2in

\centerline{\bf References}

\vskip0.1in

\noindent [1]. Gudder S. An Order for quantum observables.
\emph{Math. Slovaca}. {\bf 56}: 573-589,  (2006)

\noindent [2]. Kadison, R. Order properties of bounded self-adjoint
operators.  \emph{Proc. Amer. Math. Soc}. {\bf 34}: 505-510, (1951)

\noindent [3]. Moreland, T, Gudder, S. Infima of Hilbert space
effects. \emph{Linear Algebra and Its Applications}. {\bf 286}:
1-17, (1999)

\noindent [4]. Gudder, S. Lattice properties of quantum effects.
\emph{J. Math. Phys.}. {\bf 37}: 2637-2642, (1996)

\noindent [5]. Ando, T. Problem of infimum in the positive cone.
\emph{Analytic and Geometric Inequalities and Applications}. {\bf
478}: 1-12, (1999)

\noindent [6]. Du Hongke, Deng Chunyuan, Li Qihui. On the infimum
problem of Hilbert space effects. \emph{Science in China: Series A
Mathematics}. {\bf 49}: 545-556£¬ (2006)

\noindent [7]. Pulmannova S, Vincekova E. Remarks on the order for
quantum observables. \emph{Math Slovaca}. {\bf 57}: 589-600, (2007)

\noindent [8]. Liu Weihua, Wu Junde. A representation theorem of
infimum of bounded quantum observables. \emph{J. Math. Phys.}. {\bf
49}: 073521, (2008)

\end{document}